\documentclass[aps,prd,preprintnumbers,showpacs,superscriptaddress,nofootinbib,amsmath,amssymb,floats,floatfix,showkeys,notitlepage,longbibliography]{revtex4-1}
\usepackage[table]{xcolor} % For table row colors
\usepackage{array} % For custom column alignments
\definecolor{lightblue}{rgb}{0.62, 0.83, 0.96} % Light blue color
\definecolor{lightgray}{rgb}{0.93, 0.97, 0.99} % Light gray color
\usepackage{comment}
\usepackage{graphicx}
\usepackage{subfigure}
\usepackage{epstopdf}
\usepackage{palatino}
\usepackage[commandnameprefix=always]{changes}
\usepackage{hyperref}
\hypersetup{colorlinks=true,linkcolor=red,urlcolor=red,citecolor=red}
\usepackage[toc,page]{appendix}
\usepackage[normalem]{ulem}
\usepackage{lipsum}
\usepackage{graphicx}
\usepackage{subfigure}
\usepackage{palatino}
\usepackage{float}
\usepackage{sans}
\usepackage{adjustbox}
\usepackage{latexsym}
\usepackage{amsmath}
\usepackage{amssymb}
\usepackage{amsfonts}
\usepackage{dcolumn}
\usepackage{bm}
\usepackage{tikz}
\usepackage{bigints}
\usepackage{array,tabularx,multirow,booktabs}
\usepackage[tracking=true]{microtype}
\SetTracking{}{500}
\SetTracking{encoding={*}, shape=sc}{40}
\allowdisplaybreaks
\usepackage{adjustbox}
\usepackage{latexsym}
\usepackage{amsmath}
\usepackage{amssymb}
\usepackage{amsfonts}
\usepackage{dcolumn}
\usepackage{bm}
\usepackage{tikz}
\usepackage{bigints}
\usepackage{array,tabularx,multirow,booktabs}
\usepackage[tracking=true]{microtype}
\usepackage{color}
\allowdisplaybreaks

\begin{document}

\title{Assessing WGC Compatibility in ModMax Black Holes via Photon Spheres Analysis and WCCC Validation}

\author{Saeed Noori Gashti}
\email{saeed.noorigashti@stu.umz.ac.ir}
\affiliation{School of Physics, Damghan University, Damghan 3671645667, Iran}

\author{Mohammad Ali S. Afshar}
\email{m.a.s.afshar@gmail.com}
\affiliation{Department of Physics, Faculty of Basic
Sciences, University of Mazandaran\\ P. O. Box 47416-95447, Babolsar, Iran}
\affiliation{School of Physics, Damghan University, Damghan 3671645667, Iran}
\affiliation{Canadian Quantum Research Center, 204-3002 32 Ave Vernon, BC V1T 2L7, Canada}

\author{Mohammad Reza Alipour}
\email{mr.alipour@stu.umz.ac.ir}
\affiliation{Department of Physics, Faculty of Basic
Sciences, University of Mazandaran\\ P. O. Box 47416-95447, Babolsar, Iran}
\affiliation{School of Physics, Damghan University, Damghan 3671645667, Iran}

\author{\.{I}zzet Sakall{\i}
%\orcidlink{0000-0001-7827-9476}
}
\email{izzet.sakallı@emu.edu.tr}
\affiliation{Physics Department, Eastern Mediterranean
University, Famagusta, 99628 North Cyprus, via Mersin 10, Turkiye}

\author{Behnam Pourhassan}
%\orcidlink{0000-0003-1338-7083}}
\email{b.pourhassan@du.ac.ir}
\affiliation{School of Physics, Damghan University, Damghan 3671645667, Iran}
\affiliation{Center for Theoretical Physics, Khazar University, 41 Mehseti Street, Baku, AZ1096, Azerbaijan}
\affiliation{Centre of Research Impact and Outcome, Chitkara University, Rajpura-140401, Punjab, India.}

\author{Jafar Sadeghi}
\email{pouriya@ipm.ir}
\affiliation{Department of Physics, Faculty of Basic
Sciences, University of Mazandaran\\ P. O. Box 47416-95447, Babolsar, Iran}
\affiliation{Canadian Quantum Research Center, 204-3002 32 Ave Vernon, BC V1T 2L7, Canada}
\begin{abstract}
It seems that the regime of Hawking radiation and evaporation ultimately drives charged black holes toward super-extremality of the charge parameter and the dominance of extremal conditions. This progression, in turn, lays the groundwork for satisfying the necessary conditions for the Weak Gravity Conjecture (WGC). Preliminary studies indicate that black holes such as the Reissner–Nordström (RN) model, in their initial form, lack the capacity to sustain super-extremality of the charge parameter. If such conditions arise, these black holes transition into naked singularities—a scenario that is highly undesirable due to the loss of causality and the breakdown of space–time geometry. This raises whether the inability to sustain super-extremality is an inherent property of the model or a consequence of the approximations and precision limitations employed in its construction. To address this, we turned to the ModMax model, which represents an extension of the RN model. Our analysis revealed that the ModMax model not only accommodates super-extremality of the charge parameter but also, under certain conditions, emerges as a promising candidate for investigating the WGC. Furthermore, we independently observed how the inclusion of the de Sitter radius ($\ell$) in the AdS model and $f(R)$ gravitational corrections—both of which enhance and complicate the model—can have a direct impact on the range of super-extremal charge tolerance which, in turn, provides the realization of the conditions necessary for the WGC.
\end{abstract}

\date{$\today$}


\maketitle
\tableofcontents
\section{Introduction}\label{s1}
One of the most intriguing aspects in the study of charged black hole (BH) dynamics is the investigation of the conditions under which BHs achieve extremality and the phenomenon whereby the BH’s charge parameter becomes super-extremal. In general, when examining BHs---and given the necessary conditions for their formation---researchers are always keen to ensure that gravity serves as the key determinant and primary justification for all the behaviors expected of a BH. However, since the concept of temperature and the gradual evaporation of BHs via Hawking radiation \cite{1,1',1''} was introduced, it has become evident that, for standard BHs, this evaporation regime eventually causes the BH to self-transition towards extremal and even super-extremal conditions \cite{2,2',2'',3,3'}. As the BH approaches its extremal state---where Hawking temperature tends to zero, radiation ceases, the Cauchy and event horizons, if present, coincide, and most importantly, the charge-to-mass ratio increases beyond one---nothing remains stable or normal in this situation. Clearly, such conditions cannot persist as a stable phase for the rest of the BH's existence.\\

At these critical moments, either the concept of the BH must be abandoned as a studiable entity consistent with causality and general relativity (leading to a “Swampland” scenario for the survival of the gravitational model), or a shock is required. This necessary shock might be explained through the WGC, acting as a safeguard for causality and general relativity. Assuming the temporary dominance of electromagnetism over gravity, and using hypotheses such as Schwinger pair production, the conjecture proposes a way to separate excess charge from the BH, enabling its recovery and return to the realm of general relativity and causality. To provide a preliminary understanding of the concepts discussed in the article, a brief explanation of the WGC is given, though we strongly encourage interested readers to explore more detailed discussions in references such as \cite{2,2',2''}. A fundamental principle underlying the swampland program is the prohibition of global symmetries in quantum gravity, allowing only gauge symmetries as exceptions. In line with this principle, the WGC, which originates from this framework, serves as a benchmark for validating field theories in accordance with our empirical understanding. The WGC is founded on the premise that gravity must be the weakest force among all fundamental interactions \cite{4}. The conjecture proposes a condition on effective field theories coupled to gravity (i.e., any consistent theory that arises from the combination of gravity and electromagnetism---or other gauge forces). It states that in any consistent quantum gravity theory, there must exist at least one particle with a charge-to-mass ratio greater than or equal to that of an extremal BH. In other words, there must exist at least one charged particle for which its electromagnetic repulsion is stronger than its gravitational attraction, thereby demonstrating gravity’s relative weakness compared to electromagnetism for at least one particle.\\

The implications of this conjecture are significant: it prevents the existence of entirely and perpetually stable BHs under all conditions, which could otherwise lead to remnants that violate entropy bounds. Furthermore, the WGC ensures that BHs dominated by electromagnetism over gravity can decay, as charged particles satisfying the conjecture can be emitted \cite{2,2',2''}. For these conditions to be realized, a BH must inherently possess the capacity to manifest and sustain super-extremal charges ($q/m > 1$) within its initial structural configuration. However, preliminary studies indicate that some charged BHs---whether modeled via the simple linear electromagnetic framework of the Reissner–Nordström (RN) solution \cite{3,3',5} or via more intricate non-linear formulations---are incapable of withstanding super-extremal charge conditions in their original form. When such conditions arise, the BH tends to evolve into a naked singularity---a scenario we strongly wish to avoid, given its implication of a breakdown in causality and space–time geometry. Therefore, among the various BH models, we must search for those that not only have the inherent capacity to develop super-extremal charges in their initial configuration but also preserve the fundamental conditions required of a BH; that is, they should not violate the Weak Cosmic Censorship Conjecture (WCCC) \cite{5,6,6',6'',6''',6'''',6a,6b} and must continue to exhibit the general observational signatures of BHs, such as the presence of a photon sphere \cite{3,3',7,7',7''}.\\

It is important to emphasize that one might question whether the mere non-violation of the WCCC and the presence of features like a photon sphere are sufficient for maintaining a BH’s integrity. Note that we are not modeling new structures here. Instead, we are studying a model that, until a few steps prior, fully adhered to the general conditions of a BH. However, under the regime predicted for its evaporation, it is advancing, and even approaching, conditions of instability that could lead to its complete loss. Therefore, if, at the end of this evolutionary trajectory and proximity to instability, the model can still display simultaneous signs of life---such as non-violation of the WCCC or the presence of a photon sphere---there is reason for optimism that the condition holds. Subsequently, we extend our analysis by incorporating the de Sitter radius to assess whether introducing this parameter affects the model’s ability to withstand super-extremal charges and facilitates the establishment of the initial conditions necessary for the realization of the WGC. In the final step, we will take a further step by considering gravitational $f(R)$ corrections applied to the four-dimensional ModMax model in order to investigate whether, with such modifications, our BH model can still, even if only temporarily, exhibit conditions in which electromagnetism supersedes gravity and thereby permit the occurrence of the WGC. In conclusion, by comparing these models, we aim to ascertain how each additional parameter incorporated into the BH action influences the capacity to sustain super-extremal charges.

\section{Photon Sphere}\label{s2}
In this section, we delve into the photon sphere of BHs. To begin, we define key terms and utilize the relevant equations that characterize these BHs. Since we will deal with the static spherical model in this study, the general form of the line element is given by,
\begin{equation}\label{Ph0}
ds^2 = -f(r) dt^2 + \frac{dr^2}{f(r)} + r^2 \left( d\theta^2 + \sin^2\theta \, d\phi^2 \right),
\end{equation}
where $f(r)$ is the metric function. The horizon radius of the BH, $r_h$, is identified as the largest root of the metric function. Typically, to study null geodesics and achieve the photon sphere, an effective potential is required; due to the $\mathbb{Z}_{2}$ symmetry, it can be practically limited to the equatorial plane ($\theta=\pi/2$) without loss of generality. In general, after some calculation, it can be written in the following form \cite{ph1,ph2},
\begin{equation}\label{Ph1}
\dot{r}^2 + V_{\text{eff}} = 0.
\end{equation}
Here, the effective potential $V_{\text{eff}}$ is expressed as,
\begin{equation}\label{Ph2}
V_{\text{eff}} = g(r) \left[ \frac{L^2}{h(r)} - \frac{E^2}{f(r)} \right],
\end{equation}
where $E$ and $L$ correspond to the photon’s energy and angular momentum, associated with the Killing vector fields $\partial_t$ and $\partial_\phi$, respectively. Due to the spherical symmetry of the solution, the photon sphere occurs at a radial position $r_{\text{ps}}$, determined by the conditions:
\begin{equation}\label{Ph3}
V_{\text{eff}} = 0, \quad \partial_r V_{\text{eff}} = 0. 
\end{equation}
This leads to the equation
\begin{equation}\label{Ph4}
\left( \frac{f(r)}{h(r)} \right)' \Bigg|_{r=r_{\text{ps}}} = 0, 
\end{equation}
where the prime denotes differentiation with respect to $r$. For an unstable photon sphere, $\partial_{r,r} V_{\text{eff}}(r_{\text{ps}}) < 0$, while $\partial_{r,r} V_{\text{eff}}(r_{\text{ps}}) > 0$ corresponds to a stable photon sphere. Differentiating further yields,
\begin{equation}\label{Ph5}
f(r) h'(r) - f'(r) h(r) = 0.
\end{equation}
At the BH horizon, where $f(r_h) = 0$, the first term vanishes, but the second term is typically nonzero. Hence, $r_{\text{ps}} \neq r_h$ in general. However, for an extremal BH---where the two horizons merge---we find $f(r_h) = 0$ and $f'(r_h) = 0$, making the photon sphere coincide with the extremal BH horizon.\\ In this article, instead of using the conventional method, we study the behavior of the photon sphere based on the topological method. In this method, the behavior of each photon sphere is analyzed by considering its topological charge. Since the methodology of this method has been well described and examined in various models in previous articles \cite{ph1,ph2,ph3,ph4,ph5,ph5'}, we refrain from repeating the details and mention only the basic relationships. 
To explore the topology of the photon sphere, we introduce the regular potential function,
\begin{equation}\label{Ph6}
H(r, \theta) = \sqrt{-\frac{g_{tt}}{g_{\phi \phi}}} = \frac{1}{\sin \theta} \sqrt{\frac{f(r)}{h(r)}}. 
\end{equation}
The photon sphere radius is obtained as the root of $\partial_r H = 0$. Utilizing the vector field components $\phi = (\phi^r, \phi^\theta)$, we define
\begin{equation}\label{Ph7}
\phi^r = \frac{\partial_r H}{\sqrt{g_{rr}}} = \sqrt{g(r)} \partial_r H, \quad \phi^\theta = \frac{\partial_\theta H}{\sqrt{g_{\theta \theta}}} = \frac{\partial_\theta H}{\sqrt{h(r)}}.
\end{equation}
The vector can be rewritten as,
\begin{equation}\label{Ph8}
\phi = \|\phi\| e^{i\Theta}, \quad \|\phi\| = \sqrt{\phi^a \phi_a}.
\end{equation}
Thus, the vector field can alternatively be expressed as $\phi = \phi^r + i \phi^\theta$, and the normalized vectors are given by,
\begin{equation}\label{Ph9}
n^a = \frac{\phi^a}{\|\phi\|}, \quad \text{with } (\phi^1 = \phi^r, \, \phi^2 = \phi^\theta).
\end{equation}

\section{Model I: ModMax BH}\label{s4}
The spherically symmetric static (SSS) solution to the Einstein equations coupled with ModMax nonlinear electrodynamics (NLED), derived in \cite{m2}, is characterized by three key parameters: the BH mass $M$, the BH charge $Q$, and the nonlinear parameter $\gamma$. The line element for this solution is expressed as \cite{m2,m3},
\begin{equation}\label{1}
ds^2 = g_{\mu\nu} dx^\mu dx^\nu = -f(r) dt^2 + \frac{dr^2}{f(r)} + r^2 d\Omega^2,
\end{equation}
where the metric function $f(r)$ is given by,
\begin{equation}\label{2}
f(r) = 1 - \frac{2M}{r} + \frac{e^{-\gamma} Q^2}{r^2}.
\end{equation}
Here, the angular part of the metric is represented as,
\begin{equation}\label{3}
d\Omega^2 = d\theta^2 + \sin^2\theta \, d\phi^2.
\end{equation}
The charge $Q$ can be purely electric ($Q = Q_e$), purely magnetic ($Q = Q_m$), or a combination of both (the dyonic case), such that
\begin{equation}\label{3}
Q = \sqrt{Q_e^2 + Q_m^2}.
\end{equation}
This metric describes a charged BH with horizons determined by the roots of the equation $f(r) = 0$. These roots are given by,
\begin{equation}\label{4}
r_+ = M + \sqrt{M^2 - e^{-\gamma} Q^2}, \quad r_- = M - \sqrt{M^2 - e^{-\gamma} Q^2}.
\end{equation}
For the existence of an event horizon at $r_+$, the condition $0 \leq e^{-\gamma} Q^2 \leq M^2$ must be satisfied. The metric $g_{\mu\nu}$ in this form is regarded as the background metric for subsequent analyses. The metric function for the Reissner–Nordström BH (RN BH) is given as,
\begin{equation}\label{5}
f_{\text{RN}}(r) = 1 - \frac{2M}{r} + \frac{Q^2}{r^2}.
\end{equation}

\begin{figure}[htbp]
  \centering
  % Subfigure (a) - Schwarzschild Black Hole
  \subfigure[ Schwarzschild BH]{%
    \includegraphics[width=0.5\linewidth]{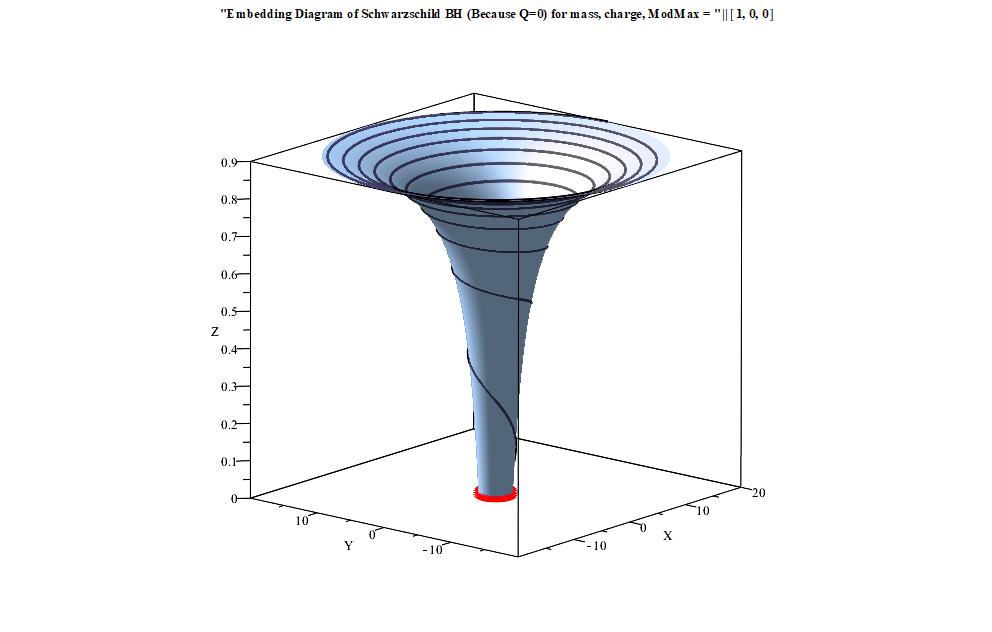}%
    \label{fig:schw}%
  }
  \hspace{0.03\linewidth}
  % Subfigure (b) - Reissner–Nordström Black Hole
  \subfigure[RN BH]{%
    \includegraphics[width=0.45\linewidth]{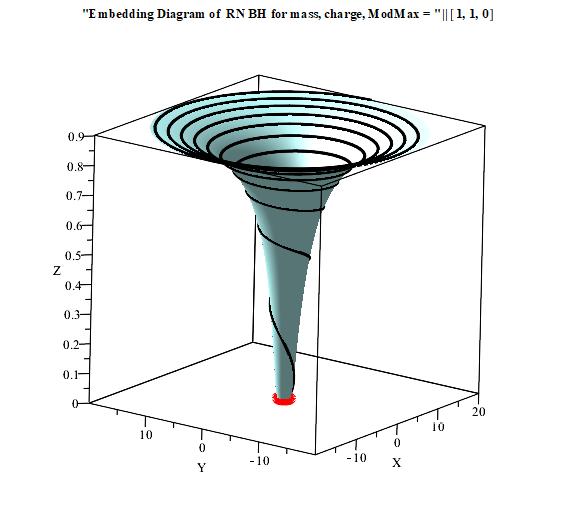}%
    \label{fig:rn}%
  }
  \hspace{0.03\linewidth}
  % Subfigure (c) - ModMax Parameter Space I
  \subfigure[ModMax Parameter Space I]{%
    \includegraphics[width=0.45\linewidth]{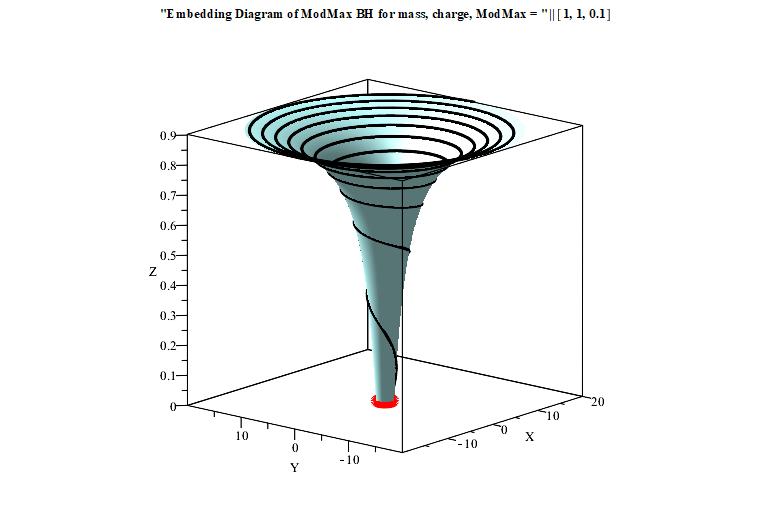}%
    \label{fig:modmax1}%
  }
  % Subfigure (d) - ModMax Parameter Space II
  \subfigure[ModMax Parameter Space II]{%
    \includegraphics[width=0.45\linewidth]{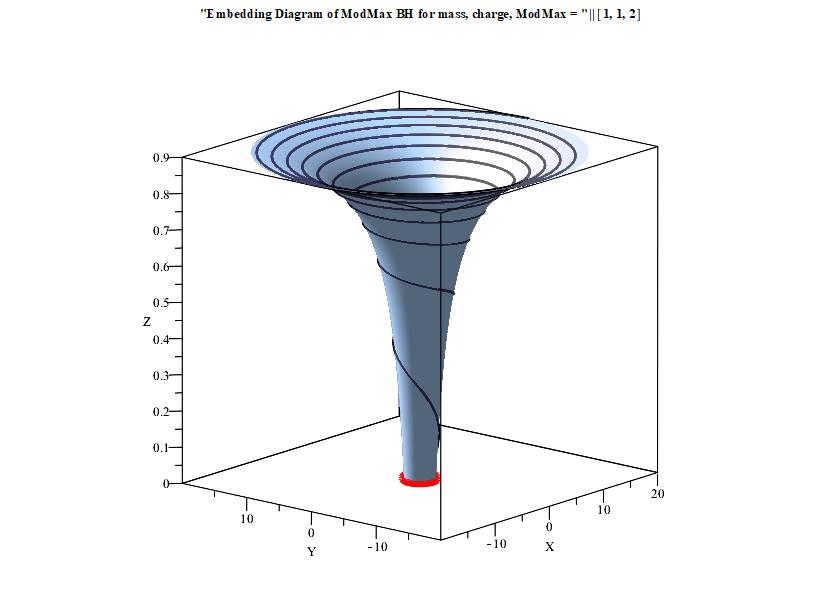}%
    \label{fig:modmax2}%
  }
  \hspace{0.03\linewidth}
  % Subfigure (e) - ModMax Parameter Space III
  \subfigure[ModMax Parameter Space III]{%
    \includegraphics[width=0.45\linewidth]{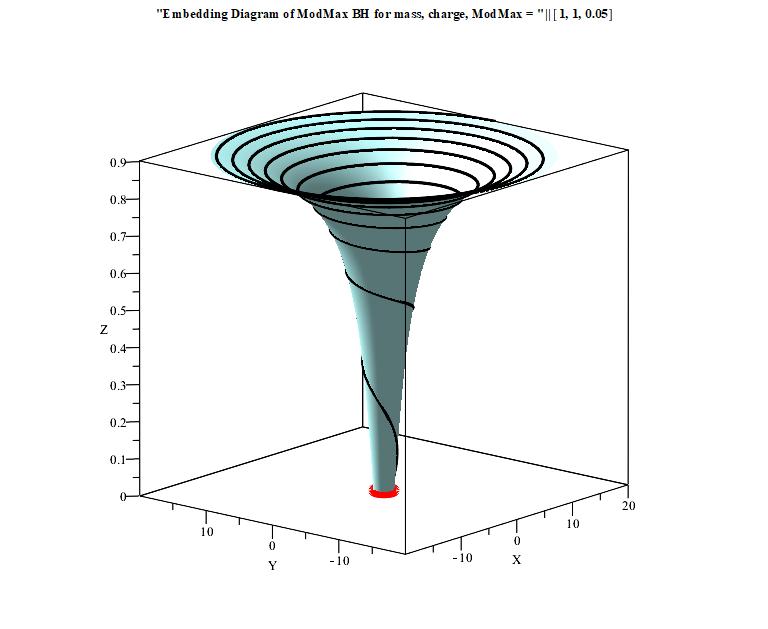}%
    \label{fig:modmax3}%
  }
  \hspace{0.03\linewidth}
  % Subfigure (f) - ModMax Parameter Space IV
  \subfigure[ModMax Parameter Space IV]{%
    \includegraphics[width=0.45\linewidth]{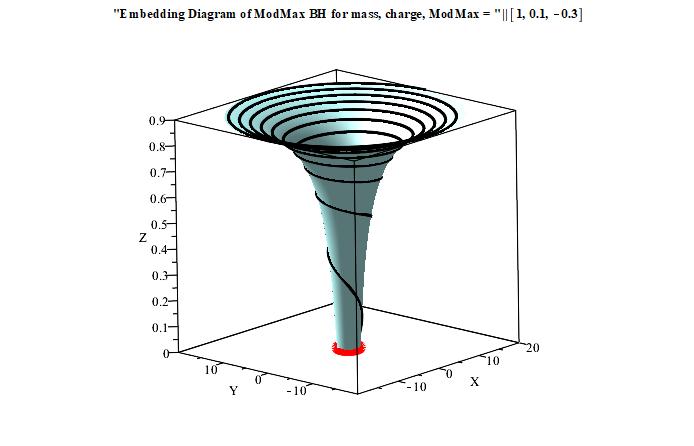}%
    \label{fig:modmax4}%
  }
 
    \label{fig:extra}%
  
  \caption{Combined view of the black hole models and related parameter spaces. (a) The Schwarzschild BH is shown for baseline comparison. (b) The RN BH demonstrates the standard charged black hole solution. (c)--(f) Different snapshots of the ModMax model illustrate various parameter regimes, including the effects of the nonlinear parameter $\gamma$ on screening and extremality conditions.}
  \label{fig:all}
\end{figure}

It is evident that the ModMax metric function $f(r)$ resembles the RN metric, but with a screened charge due to the presence of the exponential factor $e^{-\gamma}$. This screening effect allows the ModMax BH to accommodate a higher charge compared to the RN BH. Specifically:
- For the RN BH, the charge is constrained by $Q^2 \leq M^2$.
- For the ModMax BH, the charge can satisfy the less restrictive condition $Q^2 \leq e^\gamma M^2$.  In order to compare the various black hole models and their associated parameter spaces, Fig.~\ref{fig:all} presents a combined view of all relevant snapshots. Subfigure (a) displays the Schwarzschild BH for a baseline comparison, while subfigure (b) shows the RN BH, emphasizing the limitations of the linear Maxwell theory in the context of extremal conditions. Subfigures (c) through (f) reveal different aspects of the ModMax model, illustrating how the nonlinear parameter $\gamma$ influences the screening effect and how the model accommodates super-extremal charge regimes. These combined images underpin our discussion by visually contrasting the classical solutions with the modifications introduced by the ModMax model.

\begin{figure}[h!]
 \begin{center}
 \subfigure[]{
 \includegraphics[height=5cm,width=8cm]{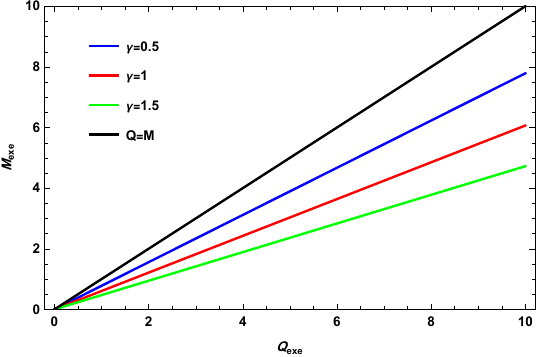}
 \label{fig5a}}
  \caption{\small{The ($Q_{exe}-M_{exe}$) plan for different values of $\gamma$.}}
 \label{fig5}
 \end{center}
 \end{figure}

\begin{figure}[h!]
 \begin{center}
 \subfigure[]{
 \includegraphics[height=4cm,width=5cm]{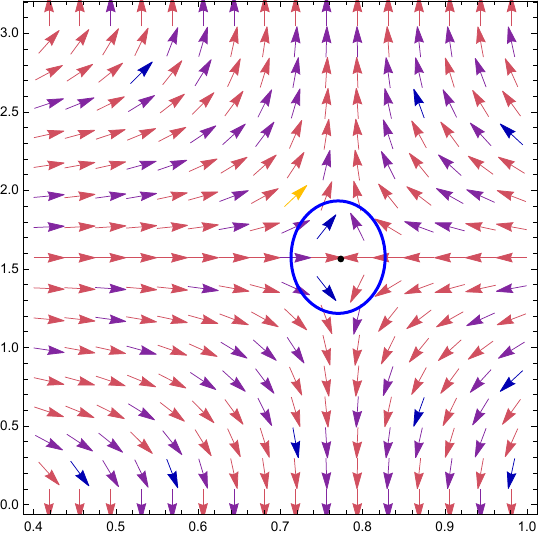}
 \label{fig6a}}
 \subfigure[]{
 \includegraphics[height=4cm,width=5cm]{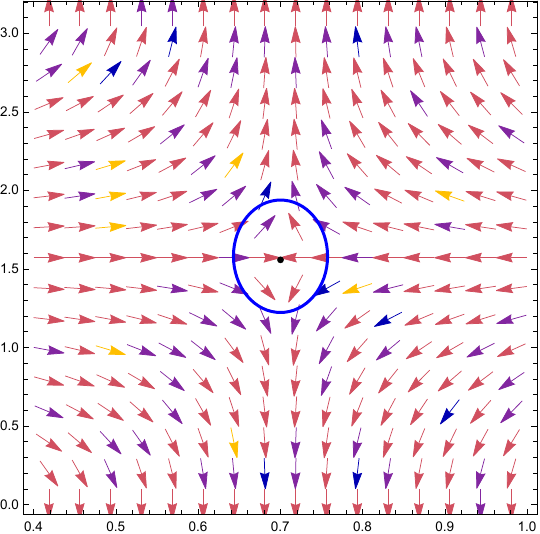}
 \label{fig6b}}
  \subfigure[]{
 \includegraphics[height=4cm,width=5cm]{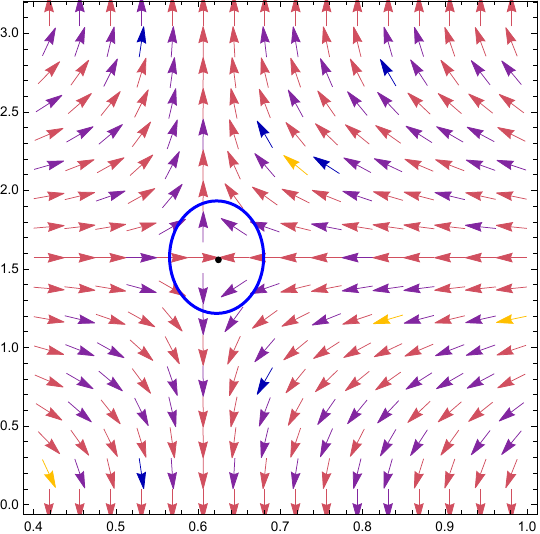}
 \label{fig6c}}
 \caption{\small{The normal vector in the $(r, \theta)$ plane associated with the photon spheres. (a) $\gamma=0.5$, $Q_{ext}=0.5$, $M_{ext}=0.3894$, $r_{ext}=0.3894$. (b) $\gamma=0.7$, $Q_{ext}=0.5$, $M_{ext}=0.35234$, $r_{ext}=0.35234$. (c) $\gamma=1$, $Q_{ext}=0.5$, $M_{ext}=0.30326$, $r_{ext}=0.30326$.}}
 \label{fig6}
 \end{center}
 \end{figure}

We can also calculate the temperature of this model as follows,
\begin{equation}\label{6}
T=\frac{1}{4 \pi}\left(\frac{1}{r}-\frac{e^{-\gamma } Q^2}{r^3}\right).
\end{equation}
When $T=0$ we have,
\begin{equation}\label{7}
r_{ext}=e^{-\frac{\gamma }{2}} Q.
\end{equation}
Therefore, with respect to Eqs. \eqref{4} and \eqref{7} we have,
\begin{equation}\label{8}
\frac{Q_{ext}}{M_{ext}}=e^{\frac{\gamma}{2}}.
\end{equation}
According to the above relationship, for $\gamma>0$, the WGC condition holds. So that for $\gamma\ll 1$, the relation in Eq. \eqref{8} can be expanded as follows,
\begin{equation}\label{9}
\frac{Q_{ext}}{M_{ext}}=1+\frac{\gamma}{2}+\mathcal{O}(\gamma).
\end{equation}
As evident from the equations and figures, Model I demonstrates structural changes with respect to the parameter $\gamma$. For all values of $\gamma > 0$, the WGC is satisfied for this structure, as clearly illustrated in Fig. (\ref{fig5}). Furthermore, Fig. (\ref{fig6}) reveals the presence of an unstable photon sphere ($PS = -1$) outside the event horizon, which serves as an initial indication of the preservation of the BH structure. Importantly, while maintaining the BH structure, the photon sphere offers intuitive support for the WGC. Simultaneously, compatibility between the two conjectures---the WGC and the WCCC---is achieved. Hence, this BH model demonstrates compatibility with the WGC.
\subsection{WGC with Photon Sphere Monitoring} 
According to Eqs. \eqref{Ph6}, \eqref{Ph7} and \eqref{2} we have,
\begin{equation}\label{10}
\begin{split}
&\phi^r=-\frac{e^{-\gamma } \csc (\theta ) \left(2 Q^2-3 e^{\gamma /2} Q r+e^{\gamma } r^2\right)}{r^4},\\[1mm]
&\phi^{\theta }=-\frac{e^{-\frac{\gamma }{2}} \cot (\theta ) \csc (\theta ) \left(Q-e^{\gamma /2} r\right)}{r^3}.
\end{split}
\end{equation} 
\section{Model II: ModMax-AdS BH}\label{sec3}
The action governing Einstein's gravity, coupled with ModMax electrodynamics and a cosmological constant in a four-dimensional spacetime, can be expressed as \cite{m1},
\begin{equation}\label{eq1}
\mathcal{I} = \frac{1}{16\pi} \int_{\partial \mathcal{M}} d^4x \sqrt{-g} \left[ R - 2\Lambda - 4\mathcal{L} \right],
\end{equation}
where $R$ is the Ricci scalar, $\Lambda$ is the cosmological constant, and $g = \det(g_{\mu\nu})$ represents the determinant of the metric tensor $g_{\mu\nu}$. For simplicity, the Newtonian gravitational constant $G$ and the speed of light $c$ are set to unity ($G = c = 1$). To include a radial electric field, we introduce the gauge potential,
\begin{equation}\label{eq3}
A_\mu = h(r) \delta^t_\mu.
\end{equation}
By utilizing the metric and the ModMax field equations,
\begin{equation}\label{eq4}
\partial_\mu \left( \sqrt{-g} e^{-\gamma} F^{\mu\nu} \right) = 0,
\end{equation}
we derive,
\begin{equation}\label{eq5}
2h'(r) + r h''(r) = 0,
\end{equation}
where the prime ($'$) and double prime ($''$) denote the first and second derivatives with respect to $r$, respectively. Solving this equation yields
\begin{equation}\label{eq6}
h(r) = -\frac{q}{r},
\end{equation}
where $q$ is an integration constant related to the electric charge. The electric field $E(r)$ can then be extracted using the electromagnetic field tensor, resulting in,
\begin{equation}\label{eq7}
E(r) = \frac{q}{r^2} e^{-\gamma}.
\end{equation}
With further calculations from \cite{m1}, the following differential equations are obtained,
\begin{equation}\label{eq8}
\text{Eq}_{tt} = \text{Eq}_{rr} = r f'(r) + \Lambda r^2 - 1 + f(r) + \frac{q^2}{r^2} e^{-\gamma} = 0,
\end{equation}
and,
\begin{equation}\label{eq9}
\text{Eq}_{\theta\theta} = \text{Eq}_{\phi\phi} = f''(r) + 2\Lambda + \frac{2}{r} f'(r) - \frac{2q^2}{r^4} e^{-\gamma} = 0.
\end{equation}
By solving these differential equations, the metric function $f(r)$ is found to be,
\begin{equation}\label{eq10}
f(r) = 1 - \frac{2m_0}{r} - \frac{\Lambda r^2}{3} + \frac{q^2 e^{-\gamma}}{r^2},
\end{equation}
where $m_0$ is an integration constant representing the geometrical mass of the BH. Notably, this solution satisfies all components of the field equations simultaneously. When $\gamma = 0$, the metric reduces to the familiar RN (A)dS BH,
\begin{equation}\label{eq11}
f(r) = 1 - \frac{2m_0}{r} - \frac{\Lambda r^2}{3} + \frac{q^2}{r^2}.
\end{equation}
Therefore, according to Eq. (\ref{eq10}), the mass and temperature for this model are calculated in the following form,
\begin{equation}\label{eq12}
M=\frac{e^{-\gamma } \left(l^2 Q^2+e^{\gamma } l^2 r^2+e^{\gamma } r^4\right)}{2 l^2 r},
\end{equation}
and,
\begin{equation}\label{eq13}
\frac{1}{4 \pi}\left(\frac{3 r}{l^2}-\frac{e^{-\gamma } Q^2}{r^3}+\frac{1}{r}\right).
\end{equation}
The WGC plays a central role within the swampland program, a framework that seeks to differentiate effective field theories (EFTs) that are compatible with quantum gravity (the “landscape”) from those that are not (the “swampland”). As a guiding principle, the WGC ensures that gravity is always the weakest force in $U(1)$ gauge theories. This condition is satisfied by the existence of particles with a charge-to-mass ratio $q/m > 1$. 
The implications of the WGC are profound when applied to BH physics. Charged BHs are generally categorized based on the relationship between their charge $Q$ and mass $M$ as follows:
- Subextremal BHs ($Q < M$): these are undercharged.
- Extremal BHs ($Q = M$): these BHs lie at the threshold of being maximally charged.
- Superextremal BHs ($Q > M$): these are overcharged.
According to the WGC, extremal BHs should decay and throw superextremal particles, thereby preventing the formation of naked singularities and maintaining consistency with the WCCC. If the WGC is violated, BHs could transition into superextremal states, exposing singularities that conflict with the WCCC and established physical laws. Consequently, the existence of superextremal particles is critical for ensuring both the stability and consistent evaporation of BHs, emphasizing their role in preserving the delicate interplay of forces described by the WGC.\\

Thermodynamics provides an essential framework for understanding BH behavior, offering insights into stability, decay mechanisms, and alignment with the WGC. Researchers utilize this framework to test the conjecture by examining cosmic phenomena, BH classifications, and theoretical predictions for observable evidence. This systematic approach not only supports the WGC's foundation but also strengthens the connection between quantum mechanics and cosmology, contributing to a more comprehensive understanding of fundamental principles.\\

To further explore the WGC and the WCCC, we study the metric of ModMax BHs in AdS spacetimes, focusing on their horizon structure and physical properties. The event horizon radius is determined by solving the condition $f(r) = 0$, where $M$ and $Q$ denote the BH's mass and charge, respectively. In RN BHs, when $Q > M$, the BH lacks an event horizon, leaving its singularity exposed to external observers as a naked singularity. This scenario directly contradicts the WCCC, which posits that singularities must always be hidden by event horizons.
By simultaneously solving the metric equation and the extremality condition (derived by setting its derivative or temperature to zero), we establish the conditions under which the WGC and the WCCC are satisfied.
Since the WGC and WCCC are not universally valid for all BHs, our investigation focuses on a specific scenario where these conjectures cannot simultaneously hold without the inclusion of the ModMax parameter. By isolating ModMax parameter-dependent terms in the metric, we identify configurations where the WCCC is satisfied alongside the WGC, at least in extremal cases. Using well-established relationships for the WGC, we calculate the extremal limit and determine the range of parameters where the WGC holds true.\\

Numerical methods are then employed to explore the regions of compatibility between the WGC and the WCCC, given the impracticality of obtaining an analytical solution for the higher-order terms in the equations. If a region exists where both conjectures are satisfied at critical or extremal points, this compatibility range can potentially extend to other areas of the BH parameter space.
By examining various BH classes and categorizing them based on their properties, we identify suitable candidates for probing quantum gravity concepts. These classifications provide a foundation for extensive research, enabling a deeper understanding of BH physics while upholding essential physical principles. Through this rigorous approach, we contribute to the broader objectives of the swampland program, reinforcing its role in bridging quantum mechanics, cosmology, and fundamental physics. For further study, you can see \cite{a,b,c,d,e,f,g,h,i,j,k,l,m,n,o,p,q,r,s,t,u,v,w,x,y,z,aa,bb,cc,dd,ee,ff,gg,hh,ii,kk,ll,mm,nn,oo,qq,rr,ss,tt,uu,vv,ww,xx,yy,zz,aaa,bbb,ccc,ddd}. To determine the extremality limit, we calculate the quantities $r_{ext}$ and $M_{ext}$ with respect to Eqs. (\ref{eq10}), (\ref{eq12}) and (\ref{eq13}). Thus, we have,
\begin{equation}\label{eq14}
r_{exe}=\frac{\sqrt{e^{-\frac{\gamma }{2}} l \sqrt{e^{\gamma } l^2+12 Q^2}-l^2}}{\sqrt{6}}.
\end{equation}
Using $r_e$ in the mass equation to determine the extremality limit, we can establish the inequality $(q^2/m^2)\geq(Q^2/M^2 )_e$,
\begin{equation}\label{eq15}
M_{exe}=\frac{e^{-\gamma } \left(-e^{\gamma } l^2+e^{\gamma /2} l \sqrt{e^{\gamma } l^2+12 Q^2}+12 Q^2\right)}{3 \sqrt{6} \sqrt{l \left(e^{-\frac{\gamma }{2}} \sqrt{e^{\gamma } l^2+12 Q^2}-l\right)}}.
\end{equation}
\subsection{WGC-WCCC Connection}
\begin{figure}[h!]
 \begin{center}
 \subfigure[]{
 \includegraphics[height=5cm,width=8cm]{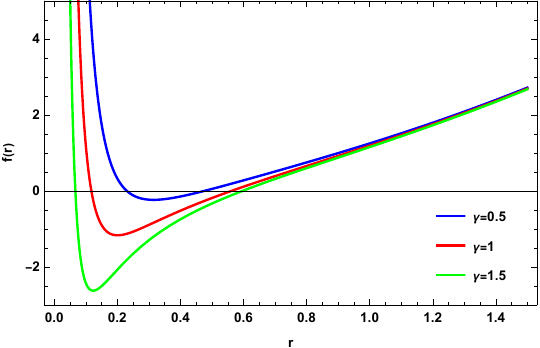}
 \label{fig1a}}
 \subfigure[]{
 \includegraphics[height=5cm,width=8cm]{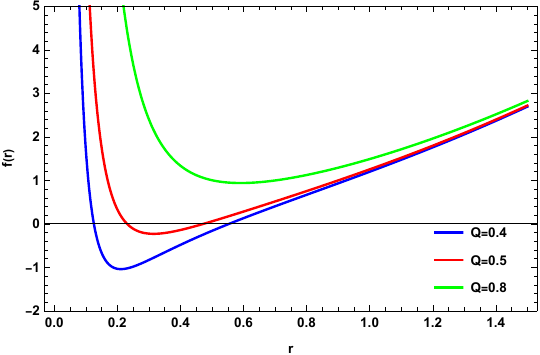}
 \label{fig1b}}
 \caption{\small{The metric function with $\ell=1$ and $M=0.45$. (a) $Q=0.5$ and various $\gamma$, (b) $\gamma=0.5$ and various $Q$.}}
 \label{fig1}
 \end{center}
 \end{figure}

Fig. (\ref{fig1}) presents the metric functions for varying free parameters, including different values of $\gamma$ and the electric charge. This visualization forms the basis for analyzing the compatibility between the WGC and the WCCC in the forthcoming calculations.
\begin{figure}[h!]
 \begin{center}
 \subfigure[]{
 \includegraphics[height=5cm,width=8cm]{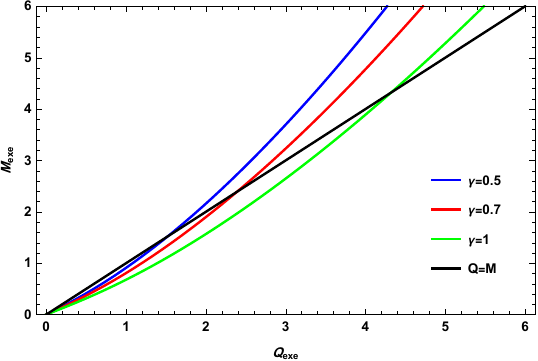}
 \label{fig2a}}
 \subfigure[]{
 \includegraphics[height=5cm,width=8cm]{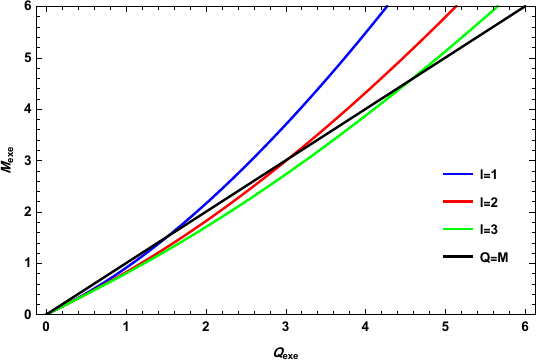}
 \label{fig2b}}
 \caption{\small{The ($Q_{exe}-M_{exe}$) plan for (a) different values of $\gamma$ with $\ell=1$, and (b) different values of $l$ with $\gamma=0.5$.}}
 \label{fig2}
 \end{center}
 \end{figure}
For Model II, structural changes influenced by the cosmological constant and the parameter $\gamma$ yield slightly different results. In this case, the WGC does not hold across all intervals, and compatibility is limited to specific ranges. At these points, the WCCC must also hold for the BH to have an event horizon.\\

As demonstrated in Fig. (\ref{fig2}), larger values of $\gamma$ correspond to a broader range of compatibility with the WGC---a result that adds significant appeal to this model.\\

\begin{figure}[h!]
 \begin{center}
 \subfigure[]{
 \includegraphics[height=4cm,width=5cm]{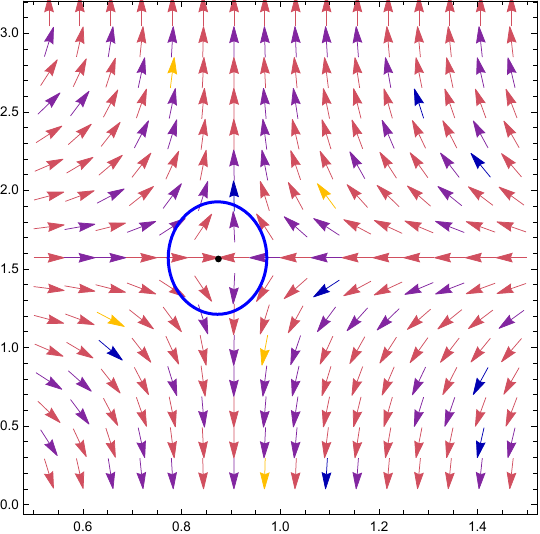}
 \label{fig3a}}
 \subfigure[]{
 \includegraphics[height=4cm,width=5cm]{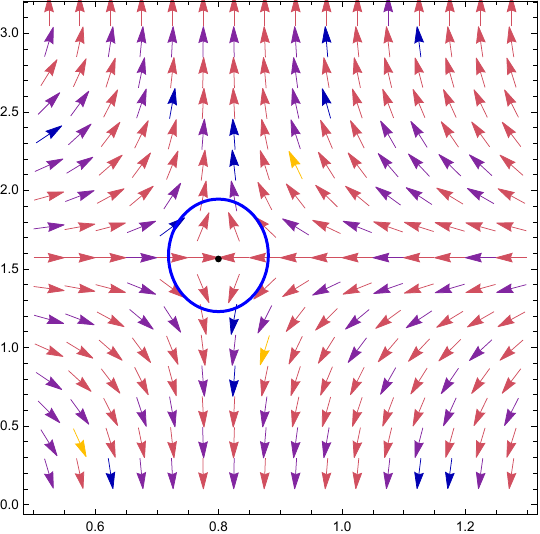}
 \label{fig3b}}
  \subfigure[]{
 \includegraphics[height=4cm,width=5cm]{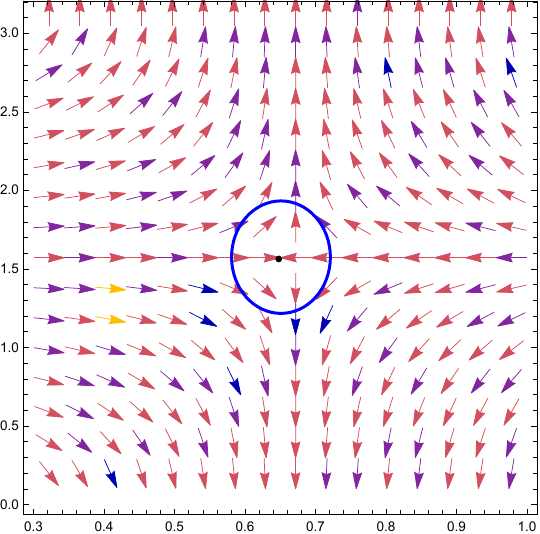}
 \label{fig3c}}
 \caption{\small{The normal vector in the $(r, \theta)$ plane associated with the photon spheres. (a) $\gamma=0.5$, $Q_{ext}=0.5$, $M_{ext}=0.412609$, $r_{ext}=0.33644$. (b) $\gamma=0.7$, $Q_{ext}=0.5$, $M_{ext}=0.3701312$, $r_{ext}=0.31035$. (c) $\gamma=1$, $Q_{ext}=0.5$, $M_{ext}=0.31511$, $r_{ext}=0.27398$ with respect to $\ell=1$.}}
 \label{fig3}
 \end{center}
 \end{figure}

\begin{figure}[h!]
 \begin{center}
 \subfigure[]{
 \includegraphics[height=4cm,width=5cm]{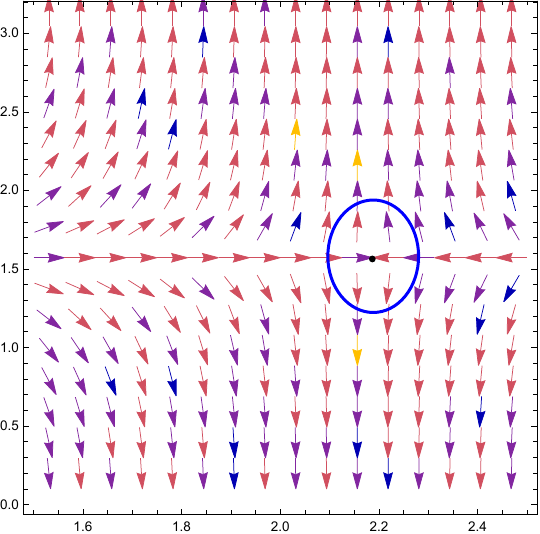}
 \label{fig4a}}
 \subfigure[]{
 \includegraphics[height=4cm,width=5cm]{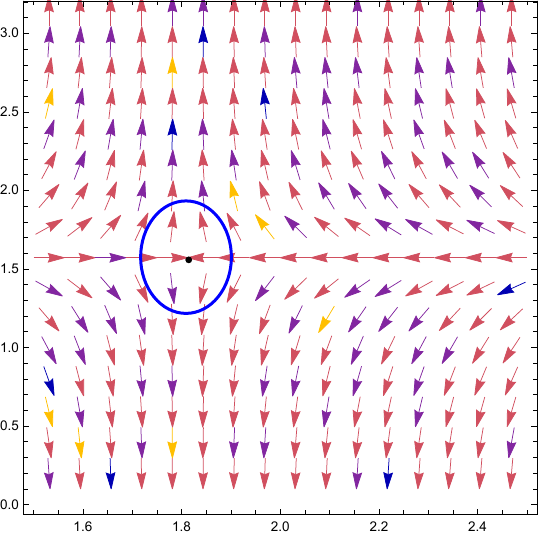}
 \label{fig4b}}
  \subfigure[]{
 \includegraphics[height=4cm,width=5cm]{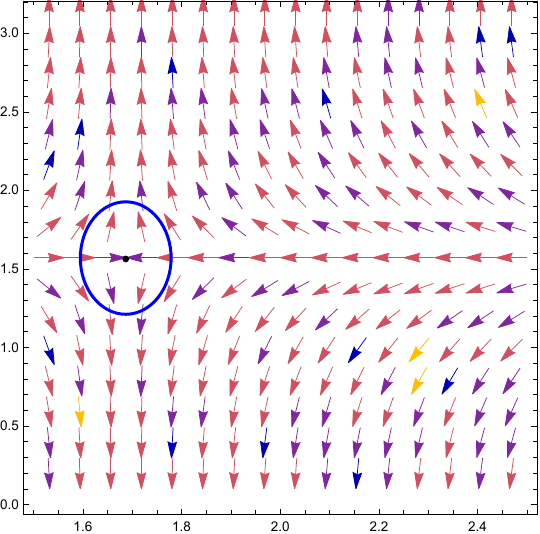}
 \label{fig4c}}
 \caption{\small{The normal vector in the $(r, \theta)$ plane associated with the photon spheres. (a) $\ell=1$, $Q_{ext}=1$, $M_{ext}=0.90935$, $r_{ext}=0.559347$. (b) $\ell=2$, $Q_{ext}=1$, $M_{ext}=0.82522$, $r_{ext}=0.672886$. (c) $\ell=3$, $Q_{ext}=1$, $M_{ext}=0.80193$, $r_{ext}=0.719252$ with respect to $\gamma=0.5$.}}
 \label{fig4}
 \end{center}
 \end{figure}

Specifically, the parameter $\gamma$ exhibits a direct relationship with the WGC, as shown in Figs. (\ref{fig3}) and (\ref{fig4}) for various parameter values. Within these ranges, an unstable photon sphere ($PS = -1$) is also observed. Additionally, the WCCC remains valid at these points, where the photon sphere provides further intuitive support for the structure as well as for the WGC.\\

Moreover, the compatibility of both conjectures is simultaneously preserved. Another parameter of particular importance is the anti-de Sitter (AdS) radius, which increases as the compatibility range for the WGC expands, as illustrated in Fig. (\ref{fig2}).
\subsection{WGC with Photon Sphere Monitoring} 
According to Eqs. (\ref{Ph6}), (\ref{Ph7}) and (\ref{eq10}) we have,
\begin{equation}\label{eq16}
\begin{split}
&\phi ^r=-\frac{\csc (\theta ) \left(r (r-3 M)+2 e^{-\gamma } Q^2\right)}{r^4},\\[1mm]
&\phi ^{\theta }=-\frac{\cot (\theta ) \csc (\theta ) \sqrt{\frac{r^2}{l^2}-\frac{2 M}{r}+\frac{e^{-\gamma } Q^2}{r^2}+1}}{r^2}.
\end{split}
\end{equation}
\newpage
\section{Model III: BH in $F(R)$-ModMax Theory }
Here, we continue our calculations with the BH solutions in $F(R)$-ModMax theory \cite{m4}. We begin by considering a static, spherically symmetric spacetime described by the line element,
\begin{equation}\label{m1}
\begin{split}
ds^2 = -g(r) \, dt^2 + \frac{dr^2}{g(r)} + r^2 \Bigl( d\theta^2 + \sin^2\theta \, d\phi^2\Bigr), 
\end{split}
\end{equation}
where $g(r)$ is the metric function to be determined. In general, the equations governing $F(R)$ gravity coupled with a nonlinear matter field (such as the ModMax field) are highly complex. As a result, obtaining an exact analytical solution can pose significant challenges. To address this, one can consider a traceless energy-momentum tensor for the nonlinear matter field. This approach allows for the derivation of an exact solution in $F(R)$ gravity coupled to the ModMax field. For a BH solution with constant scalar curvature, the trace of the stress-energy tensor $T_{\mu \nu}$ must vanish \cite{m5,m6}. By assuming the scalar curvature $R$ is constant, $R = R_0 = \text{constant}$, the trace of the field equations,
\[
R_{\mu \nu} (1 + f_R) - g_{\mu \nu} \frac{F(R)}{2} + g_{\mu \nu} \nabla^2 f_R - \nabla_\mu \nabla_\nu f_R = 8 \pi T_{\mu \nu},
\]
simplifies to,
\begin{equation}\label{m2}
R_0 \, (1 + f_{R_0}) - 2 \, (R_0 + f(R_0)) = 0, 
\end{equation}
where $f_{R_0} \equiv \frac{df}{dR} \Big|_{R=R_0}$. Solving this equation yields,
\begin{equation}\label{m3}
R_0 = \frac{2 f(R_0)}{f_{R_0} - 1}. 
\end{equation}
By substituting Eq. (\ref{m3}) into the field equations, the equations of motion in $F(R)$-ModMax theory can be expressed as,
\begin{equation}\label{m4}
R_{\mu \nu} \, (1 + f_{R_0}) - \frac{g_{\mu \nu}}{4} R_0 \, (1 + f_{R_0}) = 8 \pi T_{\mu \nu}. 
\end{equation}
It is worth noting that these equations reduce to the equations for general relativity (GR) coupled to the ModMax field when $f_{R_0} = 0$. By employing the above differential equations and assuming $R = R_0 = \text{constant}$, one can derive an exact solution. After thorough calculations, the metric function takes the form,
\begin{equation}\label{m5}
g(r) = 1 - \frac{m_0}{r} - \frac{R_0 r^2}{12} + \frac{q^2 e^{-\gamma}}{(1 + f_{R_0}) \, r^2}, 
\end{equation}
where $m_0$ is an integration constant associated with the geometric mass of the BH. Importantly, the derived solution satisfies the field equations (Eq. (\ref{m4})) under the constraint $f_{R_0} \neq -1$, ensuring physical solutions. The effects of ModMax theory are evident in the fourth term of Eq. (\ref{m5}), while the contributions of $F(R)$ gravity manifest in both the third and fourth terms. Notably, when $f_{R_0} = 0$, $R_0 = 4\Lambda$, and $\gamma = 0$, the solution reduces to the RN-(A)dS BH, with the metric function given by,
\begin{equation}\label{m6}
g(r) = 1 - \frac{m_0}{r} - \frac{\Lambda r^2}{3} + \frac{q^2}{r^2}.
\end{equation}
We analyze the conserved and thermodynamic quantities of electrically charged BHs within the framework of $F(R)$-ModMax theory. The starting point involves determining the Hawking temperature of these BHs. The Hawking temperature is computed as,
\begin{equation}\label{m7}
T = \frac{\kappa}{2\pi}, 
\end{equation}
where $\kappa$ represents the surface gravity of the BH, given by,
\begin{equation}\label{m8}
\kappa = \left((-\frac{1}{2})\nabla_\mu \chi_\nu \, \nabla^\mu \chi^\nu \right)^{1/2} = \frac{g'_{tt}}{2\sqrt{-g_{tt} g_{rr}}}  \Bigg|_{r=r_+} = \frac{g'(r)}{2} \Bigg|_{r=r_+}, 
\end{equation}
where $r_+$ is the radius of the event horizon, and $\chi = \partial_t$ is the Killing vector associated with the BH spacetime. Before obtaining the Hawking temperature, it is essential to derive an expression for the BH mass $m_0$ in terms of the event horizon radius $r_+$, the curvature parameter $R_0$, and the electric charge $q$. Solving the equation $g(r) = 0$ yields,
\begin{equation}\label{m9}
m_0 = r_+ - \frac{R_0 r_+^3}{12} + \frac{q^2 e^{-\gamma}}{(1 + f_{R_0}) r_+},
\end{equation}
where $m_0$ represents an integration constant related to the geometric mass of the BH. Using the metric function derived in Eq. (\ref{m5}) and substituting $m_0$ from Eq. (\ref{m9}) into the expression for surface gravity (Eq. (\ref{m8})), the surface gravity becomes,
\begin{equation}\label{m10}
\kappa = \frac{1}{2r_+} - \frac{R_0 r_+}{8} - \frac{q^2 e^{-\gamma}}{2 (1 + f_{R_0}) r_+^3}.
\end{equation}
Subsequently, the Hawking temperature is obtained by substituting Eq. (\ref{m10}) into Eq. (\ref{m7}), leading to,
\begin{equation}\label{m11}
T = \frac{1}{4\pi r_+} - \frac{R_0 r_+}{16\pi} - \frac{q^2 e^{-\gamma}}{4\pi (1 + f_{R_0}) r_+^3}.
\end{equation}
As shown, the Hawking temperature of BHs in $F(R)$-ModMax theory depends on various parameters, including the electric charge $q$, the curvature parameters of $F(R)$ gravity ($f_{R_0}$ and $R_0$), and the ModMax parameter $\gamma$.
The mass of the BH is calculated as
\begin{equation}\label{m12}
M=\frac{e^{-\gamma } Q^2}{2 f_{R_0} r_++2 r_+}-\frac{r_+^3 R_0}{24}+\frac{r_+}{2},
\end{equation}
and with $T=0$, the extremal radius $r_{ext}$ is calculated as follows,
\begin{equation}\label{m13}
r_{ext}=\sqrt{2} \sqrt{-\frac{\sqrt{e^{\gamma } (f_{R_0}+1) \left(e^{\gamma }+e^{\gamma } f_{R_0}+Q^2 (-R_0)\right)}}{e^{\gamma } f_{R_0} R_0+e^{\gamma } R_0}+\frac{e^{\gamma } f_{R_0}}{e^{\gamma } f_{R_0} R_0+e^{\gamma } R_0}+\frac{e^{\gamma }}{e^{\gamma } f_{R_0} R_0+e^{\gamma } R_0}}.
\end{equation}
So, we have,
\begin{equation}\label{m14}
M_e=\frac{\left(\sqrt{2} e^{-\gamma } Q\right) \left(2 e^{\gamma } (f_{R_0}+1)+\sqrt{e^{\gamma } (f_{R_0}+1) \left(e^{\gamma } (f_{R_0}+1)-Q^2 R_0\right)}\right)}{3 (f_{R_0}+1) \sqrt{e^{\gamma } (f_{R_0}+1)+\sqrt{e^{\gamma } (f_{R_0}+1) \left(e^{\gamma } (f_{R_0}+1)-Q^2 R_0\right)}}}.
\end{equation}
 \begin{figure}[h!]
 \begin{center}
 \subfigure[]{
 \includegraphics[height=5cm,width=8cm]{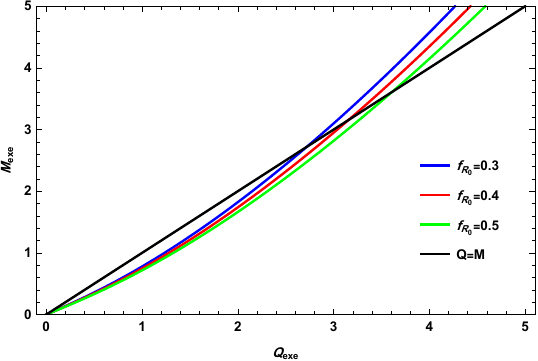}
 \label{fig7a}}
 \subfigure[]{
 \includegraphics[height=5cm,width=8cm]{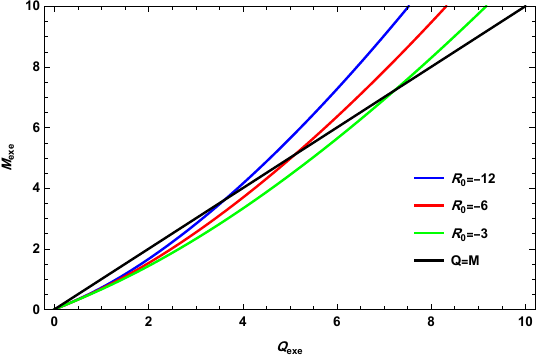}
 \label{fig7a}}
 \caption{\small{The ($Q_{exe}-M_{exe}$) plan for (a) different values of $f_{R_0}$ with $R_0=-12$ and $\gamma=0.5$, and (b) different values of $R_0$ with $f_{R_0}=0.5$ and $\gamma=0.5$.}}
 \label{fig7}
 \end{center}
 \end{figure}
\subsection{WGC with Photon Sphere Monitoring}\label{sec4}
According to Eqs. (\ref{Ph6}), (\ref{Ph7}) and (\ref{eq10}) we have,
\begin{equation}\label{eq16}
\begin{split}
&\phi^r=\frac{\csc (\theta ) \left(r (r-3 M)+\frac{2 e^{-\gamma } Q^2}{1+f_{R_0}}\right)}{r^4},\\[1mm]
&\phi^{\theta }=-\frac{\cot (\theta ) \csc (\theta ) \sqrt{\frac{e^{-\gamma } Q^2}{(1+f_{R_0}) r^2}-\frac{2 M}{r}-\frac{r^2 R_0}{12}+1}}{r^2}.
\end{split}
\end{equation}
\begin{figure}[h!]
 \begin{center}
 \subfigure[]{
 \includegraphics[height=4cm,width=5cm]{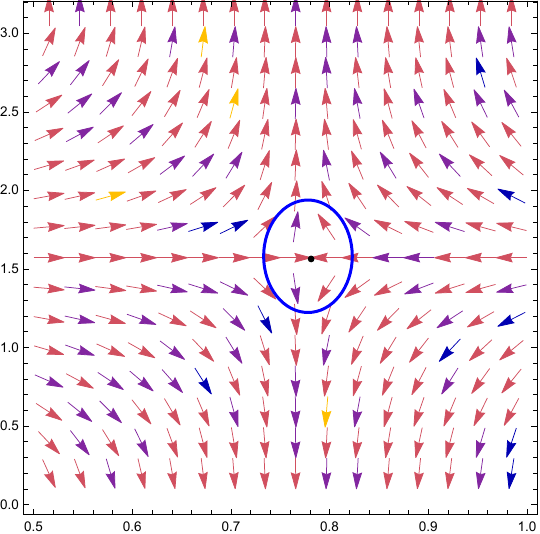}
 \label{fig8a}}
 \subfigure[]{
 \includegraphics[height=4cm,width=5cm]{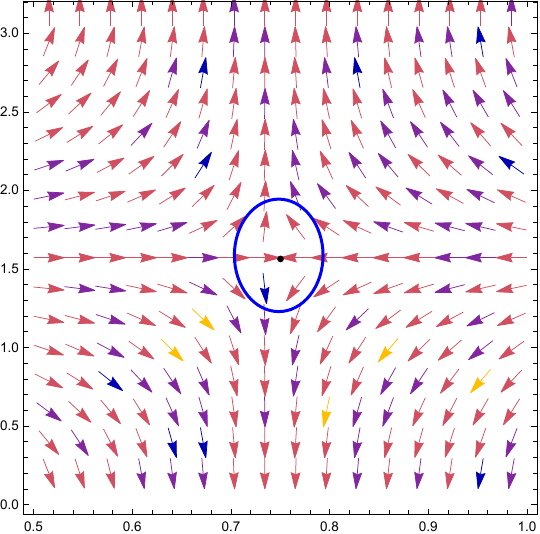}
 \label{fig8b}}
  \subfigure[]{
 \includegraphics[height=4cm,width=5cm]{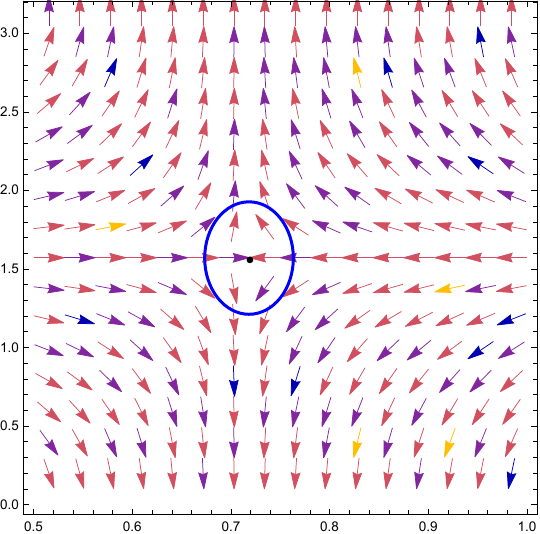}
 \label{fig8c}}
 \caption{\small{The normal vector in the $(r, \theta)$ plane associated with the photon spheres. (a) $f_{R_0}=0.3$, $Q_{ext}=0.5$, $M_{ext}=0.35788$, $r_{ext}=0.30251$. (b) $f_{R_0}=0.4$, $Q_{ext}=0.5$, $M_{ext}=0.34390$, $r_{ext}=0.29339$. (c) $f_{R_0}=0.5$, $Q_{ext}=0.5$, $M_{ext}=0.33142$, $r_{ext}=0.28508$ with respect to $\gamma=0.5$ and $R_0=-12$.}}
 \label{fig8}
 \end{center}
 \end{figure}
In Model III, additional noteworthy points emerge that distinguish it from the previous two models. For values of $R_0 > 0$, the structure reflects a de Sitter configuration, whereas for $R_0 < 0$, it adopts an anti-de Sitter (AdS) configuration. This study focuses on the AdS configuration, assuming $R_0 < 0$. In this model, the gravitational corrections play a crucial role; a larger range of the correction parameter corresponds to a higher probability of satisfying the WGC, as shown in Fig. (\ref{fig7}). Since the parameter $R_0$ is inversely related to the AdS radius, a reduction in $R_0$ enhances the compatibility range for the WGC. Additionally, Fig. (\ref{fig8}) confirms the presence of an unstable photon sphere ($PS = -1$) for the specified parameters. This provides strong intuitive support for the WGC and highlights its compatibility with the WCCC, as reflected in the diagrams.\\

To facilitate a comparative analysis of the three models with respect to the WGC, Fig. (\ref{fig9}) offers a clear depiction of the compatibility range for each model under different parameter values. Notably, Model I exhibits compatibility with the WGC for any value of $\gamma$, whereas Models II and III are subject to stricter constraints. As detailed in the text, a classification based on the degree of compatibility with the WGC can be established, ranking the models in ascending order: Model II, Model III, and Model I. This classification indicates that Model I achieves the highest level of compatibility, while Model II demonstrates the lowest.

\section{Conclusion}\label{sec5}
This study effectively investigates the critical challenge of reconciling the WGC and the WCCC within the ModMax BH framework. Through a meticulous examination of the parameter space, we identify specific conditions under which these two seemingly opposing principles achieve coherence. To substantiate this alignment, we further explore photon spheres from a topological perspective, providing valuable insights into their role in this compatibility. The topological approach, which has garnered significant attention in the scientific community, not only advances our understanding of photon spheres but also explores phase transitions in BH thermodynamics. This growing body of research underscores the importance of topology in bridging theoretical and thermodynamic aspects of BHs. For further study, you can see \cite{a19,a20,20a,21a,22a,23,24,25,26,27,28,29,31,33,34,35,37,38,38a,38b,38c,39,40,41,42,43,44,44c,44d,44e,44f,44g,44h,44i,44j,44k}.\\

As stated earlier, our motivation for studying this topic was to investigate why a basic and fundamental model such as RN BHs, when the charge parameter becomes superextremal, loses its ability to exist in the extremal form and instead transforms into a naked singularity. This phenomenon, to some extent, conflicts with the WGC, raising the question of whether this conflict stems from intrinsic properties of the Maxwell field or the degree of precision employed in formulating its electromagnetic component.\\

To address this, we considered the ModMax model, which represents a more accurate extension of Maxwellian conditions. The exponential function introduced in the ModMax model reduces to the RN model when only the first term (the least precise approximation) is retained. Our analysis demonstrates that incorporating higher-order terms of the charge in the extremal BH leads to the WGC condition being satisfied for all mass values when $\gamma$ is positive. Furthermore, this condition strengthens as $\gamma$ increases. This observation supports the hypothesis that electromagnetic models failing to exhibit conditions conducive to the emergence of the WGC may require corrections involving higher-order charge terms. In other words, this conjecture appears to possess theoretical comprehensiveness at a foundational level. However, additional observations and practical experiments are necessary for further verification. Subsequently, we examined the AdS model, which also revealed the intolerance of the RN model in this space and its inability to maintain a BH structure under superextremal charge conditions.\\

Under these circumstances, the ModMax BH satisfies the WGC condition in the extremal state but with certain restrictions. Specifically, the condition does not hold for all BH masses. Instead, for BHs with lower masses, the WGC condition persists, and with increasing $\gamma$ and AdS radius, the range of validity for the WGC expands. This extension enables heavier BHs to satisfy the conjecture. Additionally, curvature corrections, often framed as $f(R)$ gravity, have been proposed to address theoretical weaknesses in BH models---such as naked singularities and the accelerated expansion of the universe---without invoking dark energy. By replacing curvature with a generalized function, $f(R)$ gravity offers solutions involving higher-order field equations. Our study aimed to evaluate whether the WGC remains applicable under this correction, specifically within the corrected ModMax model.\\

\begin{figure}[h!]
 \begin{center}
 \subfigure[]{
 \includegraphics[height=7cm,width=12cm]{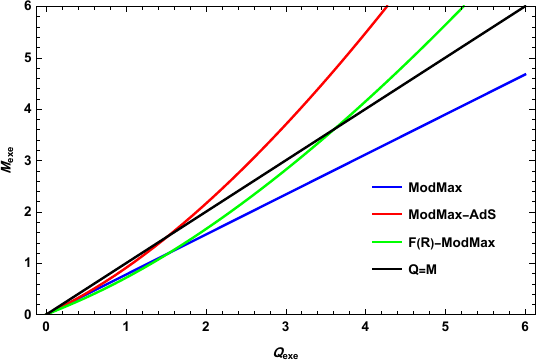}
 \label{fig9a}}
 \caption{\small{The ($Q_{exe}-M_{exe}$) plan for different ModMax BHs with $R_0=-\frac{12}{l^2}$, $l=1$, $R_0=-12$, $f_{R_0}=0.5$ and $\gamma=0.5$.}}
 \label{fig9}
 \end{center}
 \end{figure}

Our findings suggest that the WGC condition can indeed be satisfied for certain mass ranges of this BH, with the validity range increasing as $f(R)$ corrections intensify, accommodating BHs with greater rigidity. Finally, we compared the results for these three BH models in the extremal state, as illustrated in Fig. (\ref{fig9}). Based on these observations, all three studied models demonstrate initial conditions qualifying them as candidates for WGC validation.\\

The pure ModMax model stands out by encompassing a broader range of BH masses, including both lighter and heavier BHs. Conversely, the gravitational correction $f(R)$ and the AdS models introduce constraints that limit the conjecture's applicability to lighter BHs. However, greater corrections and higher $\gamma$ values extend the conditions to include heavier BHs. Despite the theoretical differences between the AdS and corrected models, a comparison can be drawn by considering the condition $(R_0=-\frac{12}{l^2})$. As shown in Fig. (\ref{fig9}), gravitational corrections offer more compatible conditions for WGC validation.
 \newpage
\section{Acknowledgements}
The work of Saeed Noori Gashti is supported by the Iran National Science Foundation (INSF). This work is based upon research funded by the INSF under project No.4038260. \.{I}.~S thanks EMU, T\"{U}B\.{I}TAK, ANKOS, and SCOAP3 for academic and/or financial support. He also acknowledges the networking support from COST Actions CA22113, CA21106, and CA23130.

\end{document}